\documentclass[prb,aps,twocolumn,preprintnumbers,amsmath,amssymb,superscriptaddress,showpacs]{revtex4}

\usepackage{amsmath}
\usepackage{graphicx}
\usepackage{amssymb}
\usepackage{dsfont}
\usepackage{color}
\usepackage[bookmarks=true,colorlinks,linkcolor=blue,urlcolor=blue,citecolor=blue]{hyperref}

\newcommand{\be}{\begin{equation}}
\newcommand{\ee}{\end{equation}}
\newcommand{\bea}{\begin{eqnarray}}
\newcommand{\eea}{\end{eqnarray}}
\newcommand{\mc}{\mathcal}

\begin{document}
\title{Chiral Spin Liquids in Arrays of Spin Chains}
\author{Gregory Gorohovsky}
\affiliation{Raymond and Beverly Sackler School of Physics and Astronomy, Tel-Aviv University, Tel Aviv, 69978, Israel}
\author{Rodrigo G. Pereira}
\affiliation{Instituto de F\'{i}sica de S\~ao Carlos, Universidade de S\~ao Paulo, C.P. 369, S\~ao Carlos, SP, 13560-970, Brazil}
\author{Eran Sela}
\affiliation{Raymond and Beverly Sackler School of Physics and Astronomy, Tel-Aviv University, Tel Aviv, 69978, Israel}

\begin{abstract}
We describe a coupled-chain construction for chiral spin liquids in two-dimensional spin systems.  Starting from a one-dimensional zigzag spin chain and imposing SU(2) symmetry in the framework of non-Abelian bosonization, we first show that our approach  faithfully describes the low-energy physics of an exactly solvable model with a   three-spin interaction. Generalizing the construction to the two-dimensional case, we obtain a theory that incorporates the universal properties of the chiral spin liquid predicted by Kalmeyer and Laughlin: charge-neutral edge states,  gapped  spin-1/2 bulk excitations, and    ground state degeneracy  on the torus signalling the  topological order of this quantum state. In addition, we show that the chiral spin liquid phase is more easily stabilized in frustrated lattices containing  corner-sharing triangles, such as the extended kagome lattice, than in the triangular lattice. Our field theoretical approach invites generalizations to more exotic chiral spin liquids and may be used to assess the existence of the chiral spin liquid as the ground state of specific lattice systems.
\end{abstract}

\maketitle

\section{Introduction}
Understanding the ground states of frustrated quantum spin systems---in which the local energetic constraints cannot all be simultaneously satisfied---is a fascinating topic in condensed matter physics~\cite{balentsNature}. One of the central proposed ground states is   Anderson's resonating valence bond state~\cite{Anderson}, a collective spin singlet  not breaking any symmetry and possessing neutral spin-$1/2$ excitations. This idea opened the way for topological phases with fractionalized excitations emerging in frustrated spin systems~\cite{Rokhsar,Read,Moessner,Senthil,Motrunich,BalentsFisherGirvin,Kitaev,Lee08,Lee06}. In 1987, Kalmeyer and Laughlin~\cite{Kalmeyer} proposed   a different spin singlet state in the triangular Heisenberg antiferromagnet that  breaks time reversal and parity symmetries, called the chiral spin liquid (CSL). In 1989, Wen, Zee and Wilczek~\cite{Wen}, and also Baskaran~\cite{Baskaran}, proposed to use the expectation value of the ``spin chirality operator'' $\mathbf{S}_i \cdot (\mathbf{S}_j \times \mathbf{S}_k)$, where $i,j,k$ belong to an elementary triangle, as an order parameter for CSLs.  Despite preserving spin SU(2) symmetry, the Kalmeyer-Laughlin CSL   shares basic properties of quantum Hall states, such as  a bulk gap and chiral edge states~\cite{Kalmeyer,fradkin,YangGirvin}.

 While it was shown later that the CSL is \emph{not }realized in the  Heisenberg antiferromagnet on the triangular lattice, a few models have been  proposed~\cite{Schroeter,Yao,Nielsen12,thomale}
 for which the CSL state is an exact ground state. However, the question remained as to whether the CSL can be realized in more realistic spin models. Recently, along with related implementations using ultracold atoms in optical lattices~\cite{NIELSEN2},  Bauer \emph{et al.}~\cite{Bauer} studied  a model of a Mott insulator on the kagome lattice  using exact diagonalization and density matrix renormalization group (DMRG) and found unambiguous evidence for realization of the Kalmeyer-Laughlin CSL. The model explicitly includes  the three-spin interaction $\mathbf{S}_i \cdot (\mathbf{S}_j \times \mathbf{S}_k)$, which is generated by an applied magnetic field. Furthermore, He, Sheng and Chen~\cite{hesheng}, as well as Gong, Zhu, and Sheng \cite{Gong}, reported a numerical  observation of a
CSL in an extended spin-$1/2$ kagomme Heisenberg model including up to next-next-nearest-neighbor interactions. The observation was again based on DMRG simulations on cylinder geometries, but in this case the spin chirality order emerged from spontaneous   breaking of time reversal symmetry.  Remarkably,  when second and third-neighbor couplings are small, instead of a CSL one finds~\cite{Gong,GongZhuBalents}  a gapped Z$_2$ spin liquid which had been identified in previous studies~\cite{YanHuseWhite}.   Quite recently, variational Monte Carlo  results have confirmed that the CSL state is energetically  favored in a  large region of the phase diagram of the extended kagome lattice  and has significant  overlap with the exact ground state obtained by exact diagonalization~\cite{HuZhuBecca,Wietek}.

In order to determine whether the ground state of a specific lattice model is a CSL, nonperturbative approaches  not restricted to finite systems are desirable.
Here we present a new field-theoretic approach  which captures all   universal properties of the Kalmeyer-Laughlin state, including fractional quasiparticle excitations and degeneracy on the torus. Our approach is based on the ``sliding Luttinger liquid" or ``coupled-wire approach" to the fractional quantum Hall effect (FQHE)~\cite{kane_02,teo_2014}. Similar constructions based on arrays of one-dimensional (1D) subsystems have proven  powerful   in the description of exotic quantum Hall states and non-Abelian anyons~\cite{poilblanc_87,yakovenko_91,sondhi_00,klinovaja_epjb_14,neupert_14,GANGOF11}, fractional topological insulators~\cite{sagi_14,klinovaja_14,Meng_14,Santos}, liquids of interacting anyons~\cite{Ludwig11,Gils09} and purely 1D systems~\cite{OregSelaStern,Meng14,Cornfeld15}.

We construct a two-dimensional (2D) CSL from an array of antiferromagnetic  Heisenberg spin chains. Leaving the detailed derivation to the bulk of the paper, here we   describe the construction pictorially. In the limit where the spin chains are decoupled, each chain has gapless spin wave excitations moving either to the left ($L$) or to the right ($R$). A topologically trivial gapped phase of the 2D spin system arises if an energy gap is produced due to coupling of $L$ and $R$ movers within the \emph{same} chain. On the other hand, a topologically nontrivial phase arises if the energy gap stems from coupling of the $L$ and $R$   modes  of \emph{neighbohring} chains. As  we demonstrate later, this picture implies the emergence of edge states for a geometry with open boundaries, consisting of the unpaired $L$ and $R$ modes in the spatially separated edge chains. Since  these edge states are charge neutral,  the Hall conductivity vanishes; yet,  they are able to conduct heat as well as spin currents.

The bulk Hamiltonian in the topological phase locks the $L$ and $R$  spin modes on neighbohring chains into an SU(2) symmetric spin singlet state.  The theory predicts that the elementary excitations carry spin 1/2 and are  charge neutral; these are the    quasiparticles of the CSL. Since the excited states in the lattice with an even number of sites must have integer spin,  the spin 1/2 elementary excitations are fractional and the ground state has topological order compatible  with   filling factor $\nu=1/2$ in the FQHE description of the CSL~\cite{Kalmeyer}.
As compared to the electronic FQHE, the extra spin  SU(2)  symmetry implies that quasiparticles and quasiholes---equivalent to spin-up and spin-down states---are degenerate.

The topological nature of the phase, which  accounts for its long-range entanglement~\cite{xiechen}, can be tested by placing the 2D surface on the torus and counting the ground state degeneracy~\cite{wen_book}. This degeneracy emerges when the operators that transport an elementary quasiparticle along the two non-contractible directions of the torus do not commute. In our construction these operators have a natural bosonized expression  which shows their noncommutativity and the resulting doubly degenerate ground state, again consistent with the defining properties of the $\nu=1/2$ FQHE.

The most crucial condition for the applicability of  our approach is that the coupling between $L$ and $R$ spin modes of neighbohring chains opens an energy gap.  To establish the feasibility of this condition, we  start by analyzing  a  model of a zigzag chain containing chiral three-spin   interactions, for which exact results by Frahm and R\"odenbeck~\cite{Frahm} provide direct support to our approach. This agreement  invites the extension   to frustrated 2D  lattice models, \emph{e.g.} variants of the triangular and kagome lattices, in which recent numerical calculations observed signatures of the CSL.  The field theory construction similarly opens the way for generalizations to more exotic chiral spin liquid phases beyond the Kalmeyer-Laughlin  state. We shall discuss this in the outlook section and leave a detailed study for   future work.

The paper is organized as follows. We start in Sec.~\ref{se:1D} with one spatial dimension. We first review the exactly solvable lattice model introduced by Frahm and R\"odenbeck~\cite{Frahm} and then apply non-Abelian bosonization techniques to recover its low-energy physics, forming the basis of our wire construction in the simplest context of two chains. We show that the spin chirality  operator opens only a partial gap in the spectrum, leaving out two gapless modes which are the seed of the chiral edge modes in the 2D case. The 2D construction is done in Sec.~\ref{se:2D}, where a renormalization group analysis is carried out to study the competition between the CSL and other conventional instabilities. In Sec.~\ref{se:properties} the properties of the ground state obtained in this chain construction are discussed, starting with the edge states. Then the quasiparticles and their creation operator are constructed in Sec.~\ref{se:QP}, and the algebra leading to the ground state degeneracy is described in Sec.~\ref{se:GSdegeneracy}. Finally, in Sec.~\ref{sec:conclusion} we conclude and discuss future directions.

\section{One-dimensional chiral spin liquid}
\label{se:1D}
In this section we use  field theory methods to analyze a  spin-$1/2$ zigzag model which (i) captures the physics of the CSL in one spatial dimension and (ii) is exactly solvable~\cite{Frahm}. This model corroborates  that our field theory construction of CSLs, which  will become more abstract in the next section on 2D generalizations,   can indeed  describe concrete lattice realizations.
\subsection{Spin model}
\label{sec:1Dmodel}
\begin{figure}
\centering
\includegraphics*[width=.6\columnwidth]{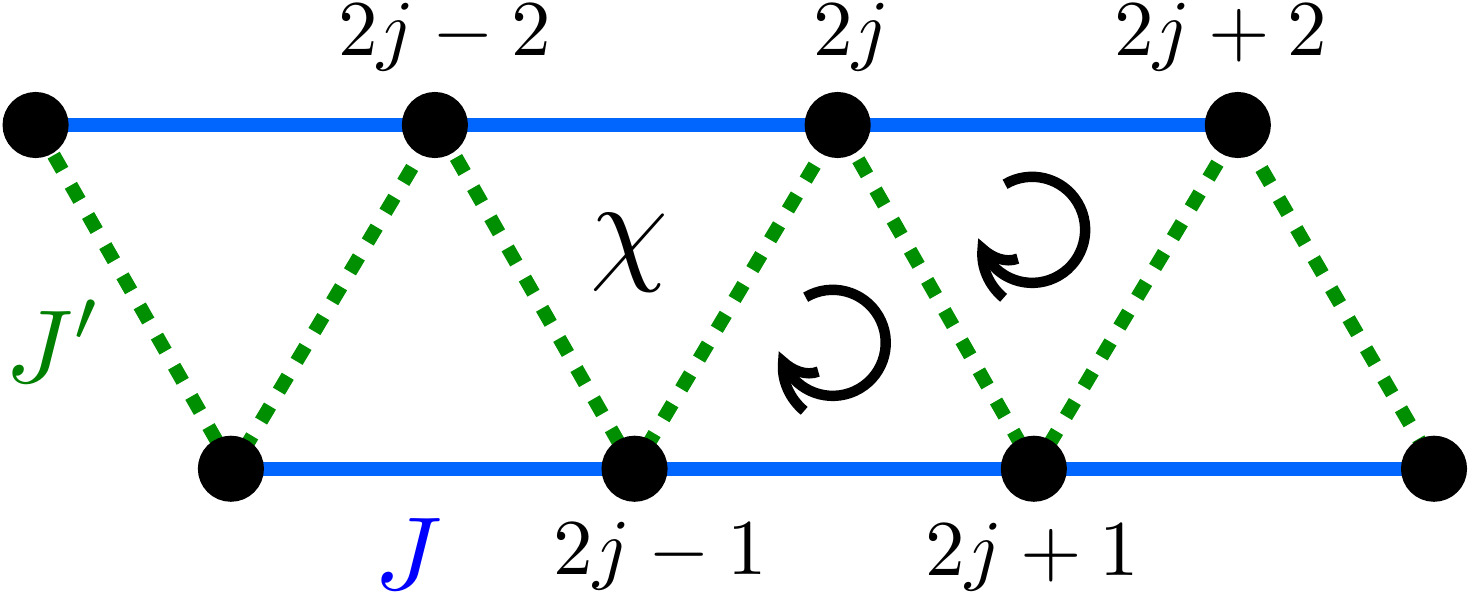}
\caption{Spin-$1/2$ zigzag chain with nearest-neighbor exchange coupling $J^\prime$, next-nearest-neighbor coupling $J$ and  three-spin interaction $\chi$. In each triangle the spins are coupled via the spin chirality operator, with the order in the triple product as indicated by the arrows (see Eq. (\ref{eq:Hchi})).}
\label{fig:zigzag}
\end{figure}

We analyze a spin-$1/2$ zigzag chain as shown in Fig.~\ref{fig:zigzag}, described by the Hamiltonian $H=H_J+H_\chi$. Here
\bea
\label{eq:H0}
H_J=\sum_j [J^\prime \mathbf{S}_{j} \cdot \mathbf{S}_{j+1}+J \mathbf{S}_{j} \cdot \mathbf{S}_{j+2}].
\eea
For dominating nearest-neighbohr antiferromagnetic exchange  $J^\prime>0$, this system behaves as a single chain perturbed by next-nearest-neighbohr coupling  $J$. The 1D Heisenberg model with $J=0$ is exactly solvable by Bethe ansatz and the ground state is in a critical phase with quasi-long-range order \cite{affleck90}. It is known that upon inclusion of a small next-nearest-neighbohr coupling $J$  the system remains critical, till an energy gap opens for $J/J^\prime \ge 0.241167$ \cite{Okamoto1992433,PhysRevB.54.R9612,affleckWhite}.
On the other hand,  in the limit  $J\gg J^\prime$ on which we will focus, the system can be thought of as two chains weakly coupled by the zigzag term $J^\prime$.
To force the system into a chiral spin state, we add  terms breaking parity and time reversal symmetry explicitly (but preserving the SU(2) symmetry)
\bea
H_{\chi} &=& \frac{\chi}{2} \sum_{j} [\mathbf{S}_{2j} \cdot (\mathbf{S}_{2j+1} \times \mathbf{S}_{2j-1})\nonumber\\
&&+\mathbf{S}_{2j+1} \cdot (\mathbf{S}_{2j} \times \mathbf{S}_{2j+2})].\label{eq:Hchi}
\eea
In both terms in Eq. (\ref{eq:Hchi}) the   spin operators appear  clockwise in the triple product (as read from left to right) with respect to every elementary triangle in Fig.~\ref{fig:zigzag}. Thus, this interaction favours uniform spin chirality. Note that   the system is not invariant under translation by one site $j \to j+1$, but only under $j \to j+2$.

It is worth mentioning that such chiral three-spin interactions  arise naturally in the Hubbard model in the presence of a magnetic flux. Following Ref.~[\onlinecite{Bauer}], consider spin-$1/2$ electrons hopping on the same zigzag lattice with nearest-neighbor hopping amplitude $t_1$  and next-nearest-neighbohr hopping $t_2$. At half filling and for strong on-site Hubbbard interaction $U \gg t_1, t_2$, we obtain the usual exchange  coupling in Eq.~(\ref{eq:H0}) with $J^\prime=4t_1^2/U$, $J=4 t_2^2/U$. Adding a magnetic  flux $0<\Phi<\pi$ through each  triangle breaks time reversal symmetry and gives rise  to an interaction involving  the spin chirality operator as in Eq.~(\ref{eq:Hchi}), with $\chi \sim \frac{ t_1^2 t_2 }{U^2} \sin (\Phi)$. Higher orders in $t/U$, which are required to describe   weak Mott insulators,  tends to enhance the ratio $\chi/J^\prime$~\cite{Bauer}.

\subsection{Exact spin-wave spectrum\label{sec:exact}}
\begin{figure}[h]
\centering
\includegraphics*[width=.8\columnwidth]{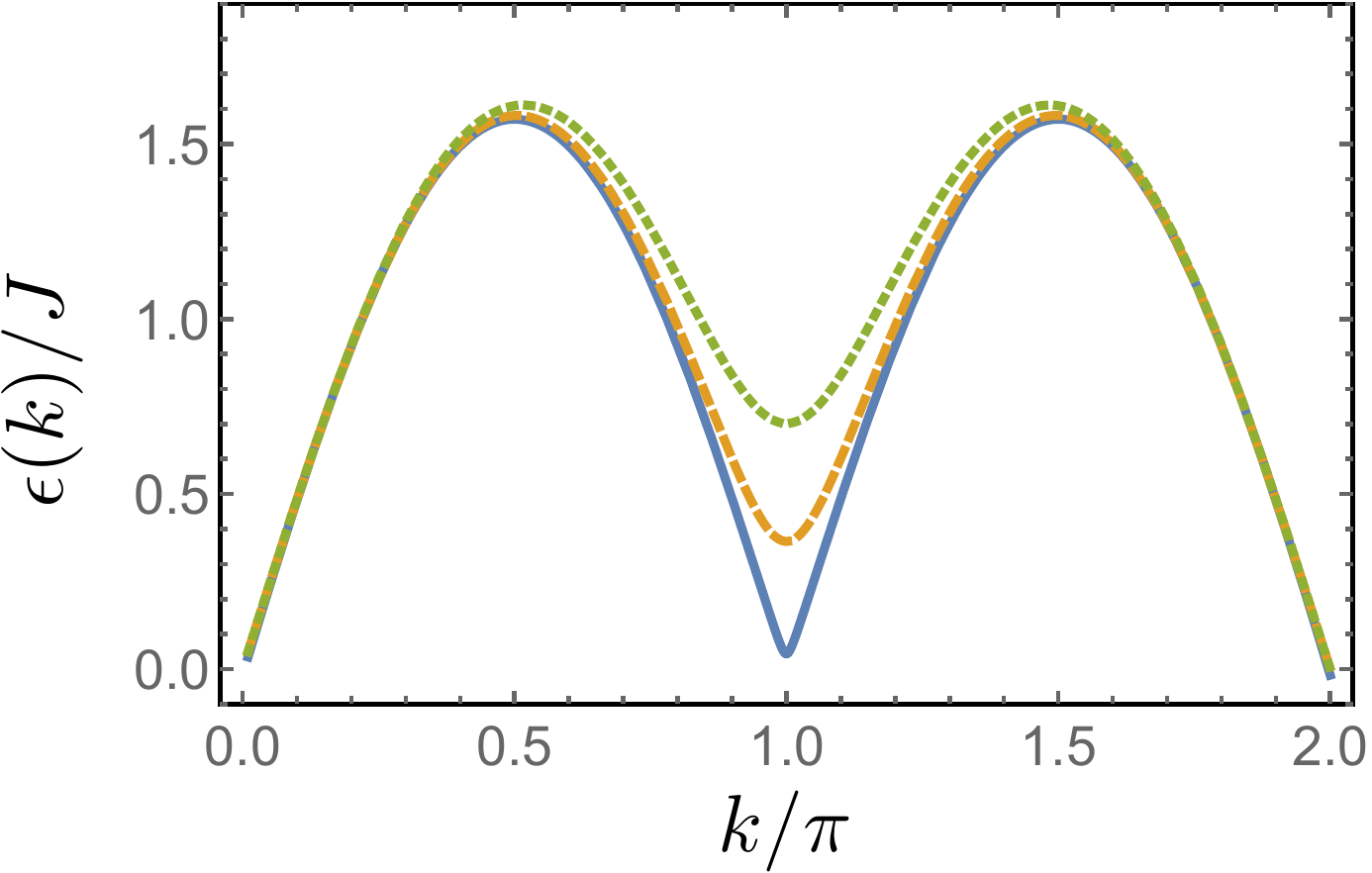}
\caption{Spinon dispersion for the exactly solvable model of Eqs.~(\ref{eq:H0}) and (\ref{eq:Hchi}) with couplings constrained as in Eq. (\ref{eq:kappa}). The three curves correspond to different values of $\kappa$; from bottom to top: $\kappa=0.99$ (solid line), $\kappa=0.97$ (dashed line), and $\kappa=0.95$ (dotted line).}
\label{fig:spinons}
\end{figure}

The Hamiltonian $H=H_J+H_\chi$ is integrable if one parametrizes the three coupling constants as~\cite{Frahm}
\be
\label{eq:kappa}
J^\prime = 2(1-\kappa),~~J=\kappa,~~ \chi=2\sqrt{\kappa(1-\kappa)}.
\ee
Varying the parameter $\kappa$ from 0 to 1 interpolates between a single Heisenberg chain and a
pair of decoupled chains.
The excitation spectrum $\epsilon(k)$ of elementary excitations --- called spinons --- has a closed-form expression extracted from the Bethe ansatz solution~\cite{Frahm}, and is plotted in Fig.~\ref{fig:spinons} for different  values of $\kappa$ near $1$.  For $\kappa = 1$ the system reduces  to two decoupled chains and one observes two  branches of excitations containing  left- ($L$) and right- ($R$) moving gapless modes at $k\text{ mod }2\pi=0$ and $k=\pi$  (in units where the lattice spacing $a=1$). For arbitrarily small deviation of $\kappa$ from unity, the pair of $R$ and $L$ movers  at $k=\pi$ acquires an energy gap\cite{Frahm}
\be
\label{Egap}
\Delta = \epsilon(k=\pi) =2 \pi J e^{-\frac{1}{\chi}},
\ee
while the pair at $k\text{ mod }2\pi=0$ remains gapless.

Below we will obtain this behavior using field theory methods. It will also be possible to explain the scaling of the energy gap with the interchain coupling. We begin by setting the notation, starting from the Hubbard model.

\subsection{Bosonization notation}
\label{sec:bosonization}
We follow the notation of Ref.~[\onlinecite{Pereira12}] and for completeness include the main formulas here. We start from the Hubbard model, where the operator $c_{j , \sigma}$ destroys an electron with spin $\sigma$  on site $j$. At long distances compared to the lattice spacing $a=1$, we expand the fermion field around the left and right Fermi points $k\approx \pm \pi/2$ and introduce chiral fermions $\psi_{L,R,\sigma}$\bea
\label{eq:c} c_{j ,\sigma} & \to & \Psi_\sigma(x) \sim e^{i \pi x /2}\psi_{R,\sigma} + e^{-i \pi x /2}\psi_{L,\sigma}.
\eea
The chiral fermions   can be subsequently bosonized as
\bea
\psi_{\alpha,\sigma}(x) &\sim &  e^{-i \sqrt{2 \pi} \varphi_{\alpha,\sigma}(x)},~~~\alpha=L,R=+,-,
\eea
where $\varphi_{\alpha,\sigma}$  are chiral bosons that obey the commutation relations
\be
\label{eq:commut}
[\varphi_{\alpha,\sigma}(x),\partial_{x'} \varphi_{\alpha',\sigma'}(x') ] =i \alpha \delta_{\alpha \alpha'} \delta_{\sigma,\sigma'} \delta(x-x') .
 \ee
 We then introduce charge and spin degrees of freedom
\bea
\varphi_{\alpha,c}(x) = \frac{\varphi_{\alpha,\uparrow}(x) + \varphi_{\alpha,\downarrow}(x)}{\sqrt{2}}  , \nonumber \\
\varphi_{\alpha,s}(x) = \frac{\varphi_{\alpha,\uparrow}(x) - \varphi_{\alpha,\downarrow}(x)}{\sqrt{2}}.
\eea
At half filling, an arbitrarily small   $U>0$ gaps out  the charge mode; this happens through the umklapp operator, whereby two electrons   of opposite spin scatter from the right  to the left Fermi point and vice versa. 
The low-energy properties are then described by the spin dynamics. From now on we will omit the spin index $s$ from the spin boson, $\varphi_{\alpha,s} \to \varphi_{\alpha}$.

The expansion of the spin operator $\mathbf{S}_j = c_{j, \sigma}^\dagger \frac{\boldsymbol{\sigma}_{\sigma \sigma'}}{2} c_{j , \sigma'}$, reads~\cite{AffleckHaldane}
\bea
\label{eq:Sexp}
\mathbf{S}_j \to \mathbf{S}(x) \sim \mathbf{J}_{R}(x) + \mathbf{J}_{L}(x) +(-1)^{x} \mathbf{n}(x).
\eea
The spin field contains two parts. The uniform part is given by   the chiral currents $\mathbf{J}_{R,L}(x)$  of the SU(2)$_1$ Wess-Zumino-Witten (WZW)  model with central charge $c=1$. In Abelian bosonization notation, \bea
\label{Jabelian}
J^z_\alpha(x)=\frac{\alpha}{\sqrt{4 \pi}} \partial_x \varphi_{\alpha}(x) ,  \nonumber \\
J^\pm_\alpha(x) =\frac{1}{2 \pi} \,e^{\pm i \sqrt{4 \pi} \varphi_{\alpha}(x)}.
\eea
The staggered part of the spin operator  can be written as~\cite{affleck90}
\be
\label{eq:ng}
\mathbf{n}(x) \propto  {\rm{tr}}[g(x) \boldsymbol{\sigma}],
\ee
where $\boldsymbol \sigma$ is the vector of Pauli matrices and
\be
\label{eq:gLR}
g(x,\tau) =  g_L(z) \otimes  g_R^\dagger(\bar{z}),
\ee
is the matrix field of the WZW model, with components  $g_{\sigma \sigma'}(x,\tau) = g^{\phantom\dagger}_{L,\sigma}(z) g^\dagger_{R,\sigma'}(\bar z)$.  Here $z = v_s \tau+i x$ and   $\bar{z} =v_s \tau-i x$ are   complex coordinates in Euclidean spacetime, with $v_s$  the velocity of the spin mode. The spinor fields $g_{L,R}$ fields have conformal dimensions~\cite{diFrancesco} $(\frac{1}{4},0)$ and  $(0,\frac{1}{4})$, respectively; in Abelian bosonization  they can be written
\be
\label{gAbelian}
g_\alpha(x) =  \left(\begin{array}{c}
      e^{-i \sqrt{\pi} \varphi_{\alpha}(x)} \\
      e^{i \sqrt{\pi} \varphi_{\alpha}(x)}
     \end{array}
    \right).
\ee

\subsection{Interchain coupling}
\label{se:fieldtheorzigzag}
We now turn to the coupling between two Heisenberg chains in the zigzag geometry and the resulting phases. This has been the subject of extensive theoretical  work, see  for example Refs.~[\onlinecite{affleckWhite,AllenSenechal,Essler}]. Here we focus on the role of the spin chirality operator $H_\chi$.

In the continuum limit for two weakly coupled chains ($J^\prime,\chi\ll J$), we write the spin operator in even ($e$) and odd ($o$) chains as
\bea
\label{eq:Snonab}
\mathbf{S}_{2i} \to  \mathbf{S}_{e}(x) \sim \mathbf{J}_{e,L}(x)+ \mathbf{J}_{e,R}(x)+(-1)^x \mathbf{n}_{e}(x),\nonumber \\
\mathbf{S}_{2i+1} \to  \mathbf{S}_{o}(x) \sim \mathbf{J}_{o,L}(x)+ \mathbf{J}_{o,R}(x)+(-1)^x \mathbf{n}_{o}(x),
\eea
and write the free  Hamiltonian  in Sugawara form \be
H_0=\sum_{l=e,o}\frac{2\pi v_s}{3}\int dx\, (\mathbf J_{l,R}^2+ \mathbf J_{l,L}^2).\label{sugawara}
    \ee
The Hamiltonian in Eq. (\ref{sugawara})   describes two pairs of gapless right-  and left-moving bosonic fields, each pair propagating in one chain.

To analyze the perturbations to Hamiltonian   (\ref{sugawara}), let us discuss the operator content of the theory and the symmetries of the lattice model.  All local operators in the WZW model can be expressed in terms of the dimension-1 chiral currents $\mathbf J_{l,\alpha}(x)$, the dimension-1/2 staggered magnetization  $\mathbf n_l(x)$ and the dimension-1/2 (SU(2) scalar) dimerization operator $\varepsilon_l(x)\propto \text{tr}[g_l(x)]$ \cite{AllenSenechal,Starykhkagomme}. These operators transform under translation $x\to x+1$ (\emph{i.e.} $j\to j+2$ in the zigzag chain)  in the form\be
\mc L:\, \mathbf J_{l,\alpha}\to   \mathbf J_{l,\alpha},\quad
\mathbf n_{l}\to  -\mathbf n_{l},\quad
\varepsilon_l\to -\varepsilon_l.
\ee
Time reversal $\mc T$ acts as follows:
\be
\mc T:\,\mathbf J_{l,R}\leftrightarrow  -\mathbf J_{l,L},\quad
\mathbf n_{l}\to  -\mathbf n_{l},\quad
\varepsilon_l\to\varepsilon_l.
\ee
Reflection $\mc P$ about an axis perpendicular to the chains that goes through an even site (site parity for the even chain and link parity for the odd chain) takes $x\to -x$ and\bea
\mc P:& \mathbf J_{l,R}\leftrightarrow  \mathbf J_{l,L},\quad
\mathbf n_{e}\to  \mathbf n_{e},\quad\mathbf n_{o}\to - \mathbf n_{o},\nonumber\\
&\varepsilon_e\to-\varepsilon_e,\quad \varepsilon_o\to\varepsilon_o.\label{parity}
\eea

In the absence of the chiral three-spin interaction  ($\chi=0$), the interchain couplings must respect  $\mc L$, $\mc P $ and $\mc T$ symmetries, as well as SU(2) invariance. It is known\cite{AllenSenechal,Starykh,Essler} that in this case the leading perturbations to Eq. (\ref{sugawara})  are  all marginal operators. First,  even the decoupled-chain Hamiltonian is perturbed by the  marginal ``backscattering''\cite{Starykh} operator
\be
\delta H_{\text{bs}}=2\pi v_s \gamma_{\text{bs}} \sum_{l=e,o}\int dx\, \mathbf J_{l,L}\cdot \mathbf J_{l,R},
\ee
with $\gamma_{\text{bs}}<0$; in addition $\gamma_{\text{bs}} = \mathcal{O}(1)$ since it stems from  the intrachain exchange coupling $J$. Second, there is the interchain current coupling \be
 \delta H_{g}  = 2\pi v_s g  \int dx   (\mathbf{J}_{e,R} \cdot \mathbf{J}_{o,L}   +\mathbf{J}_{o,R} \cdot \mathbf{J}_{e,L} ),
\ee
where $g\sim \mc O( J^\prime/J)$ is a dimensionless coupling constant. Finally, there is the ``twist'' operator \cite{Starykh,Essler}
\be
\delta H_{\text{tw}}=2\pi v_s \gamma_{\text{tw}}\int dx\,\mathbf n_e\cdot \partial_x \mathbf{n}_o,
\ee
which  carries nonzero conformal spin. The dimensionless coupling constant is $ \gamma_{\text{tw}}\sim \mc O(J^\prime/J)$.
One can then   analyze the renormalization group (RG) flow of the marginal coupling constants.
Here we have neglected the marginal current-current coupling of the form $\mathbf J_{e,R}\cdot \mathbf J_{o,R}+(R\to L)$, which does not renormalize to one-loop order \cite{affleckWhite}.
For a single Heisenberg chain,  the intrachain coupling  $\gamma_{bs}<0$  is marginally irrelevant. On the other hand, for  antiferromagnetic interchain coupling $J^\prime>0$,  both $g$ and $\gamma_{\text{tw}}$ flow to strong coupling, but $g$ reaches strong coupling first \cite{Essler}.  In this case the zigzag chain is in a  topologically trivial dimerized phase in which both pairs of  right and left movers within each chain are gapped out.

To recover the  spectrum discussed in Sec.  \ref{sec:exact}, we now consider the effects of a nonzero three-spin interaction. In this case we must allow for perturbations that are odd under $\mc P$ and $\mc T$, but invariant under the product $\mc P\circ \mc T$. We find that there is only one new marginal perturbation to the decoupled-chain Hamiltonian (and still no relevant perturbations). The latter can be obtained by taking the continuum limit in the spin chirality operator in Eq. (\ref{eq:Hchi}):
\bea
\label{eq:1}
H_{\chi} &\sim&  \frac{\chi}{2} \int dx\, \{\mathbf{S}_{e}(x)  \cdot [ \mathbf{S}_o(x+1) \times \mathbf{S}_{o}(x) ] \nonumber \\
&&+\mathbf{S}_{o}(x) \cdot [\mathbf{S}_{e}(x) \times \mathbf{S}_{e}(x+1)]\}.
\eea
We now substitute the mode  expansion Eq.~(\ref{eq:Snonab})   into $H_\chi$.
Since we have two field operators appearing at nearby positions, we must take their operator product expansion (OPE)~\cite{cardy1996scaling} on the same chain.
We can use the OPE for chiral currents \cite{AllenSenechal}
   \bea
   \label{eq:OPEJJ}
J_{l,L}^a(z) J_{l,L}^b(0)&\sim &\frac{\delta^{ab}}{8\pi^2z^2}+\frac{i}{2\pi z}\epsilon^{abc}J_{l,L}^c(0)+\dots, \nonumber \\
J_{l,R}^a(\bar z) J_{l,R}^b(0)&\sim &\frac{\delta^{ab}}{8\pi^2\bar z^2}+\frac{i}{2\pi \bar z}\epsilon^{abc}J_{l,R}^c(0)+\dots,
\eea
where $\epsilon^{abc}$ is the Levi-Civita antisymmetric tensor.  This leads to
    \bea
  \mathbf{S}_l (x+1)  \times \mathbf{S}_l (x) \sim \frac{1}{ \pi } [\mathbf{J}_{l,L}(x)-\mathbf{J}_{l,R}(x)]+...
   \eea
   as the  leading term. Keeping only nonoscillating terms in Eq.~(\ref{eq:1}), we obtain the marginal perturbation
\bea
\label{eq:Hchi1}
\delta H_{\chi} &= & 2 \pi v_s \tilde{\chi}   \int dx\,  (\mathbf{J}_{e,R} \cdot \mathbf{J}_{o,L}   -\mathbf{J}_{e,L} \cdot \mathbf{J}_{o,R} ),
\eea
with dimensionless coupling constant $\tilde{\chi} \sim \mc O( \chi/J)$. As expected, the operator in Eq. (\ref{eq:Hchi1}) is odd under both $\mc P$ and $\mc T$. We remark that we have also considered the OPE of the $\mathbf n(x)$ field with itself and with the chiral currents, but  found  no additional perturbations at the level of marginal operators.

Using the OPEs \cite{AllenSenechal,Egger} among  the fields $\mathbf J_{l,\alpha}$, $\mathbf n_l$ and $\varepsilon_l$ , we derive a set of coupled RG equations for the marginal coupling constants in the presence of the chirality operator:
\bea
\label{eq:rg4} \frac{d\gamma_{\text{bs}}}{d\ell}&=&\gamma_{\text{bs}}^2,\\
\label{eq:rg1}\frac{dg}{d\ell}&=&g^2+\tilde{\chi}^2+2\pi ^2C^2\gamma_{\text{tw}}^2,\\
\label{eq:rg2}\frac{d \tilde{\chi}}{d\ell}&=&2g \tilde{\chi},\\
\label{eq:rg3}\frac{d\gamma_{\text{tw}}}{d\ell}&=&(g+\gamma_{\text{bs}}) \gamma_{\text{tw}},
\eea
where $C$ is a nonuniversal prefactor of order unity appearing in the OPE of $\mathbf n(x)$ and  $d \ell = d \ln (\Lambda_0/\Lambda)$ with $\Lambda$ the ultraviolet momentum cutoff.

\begin{figure}
  \includegraphics[width=0.8\columnwidth]{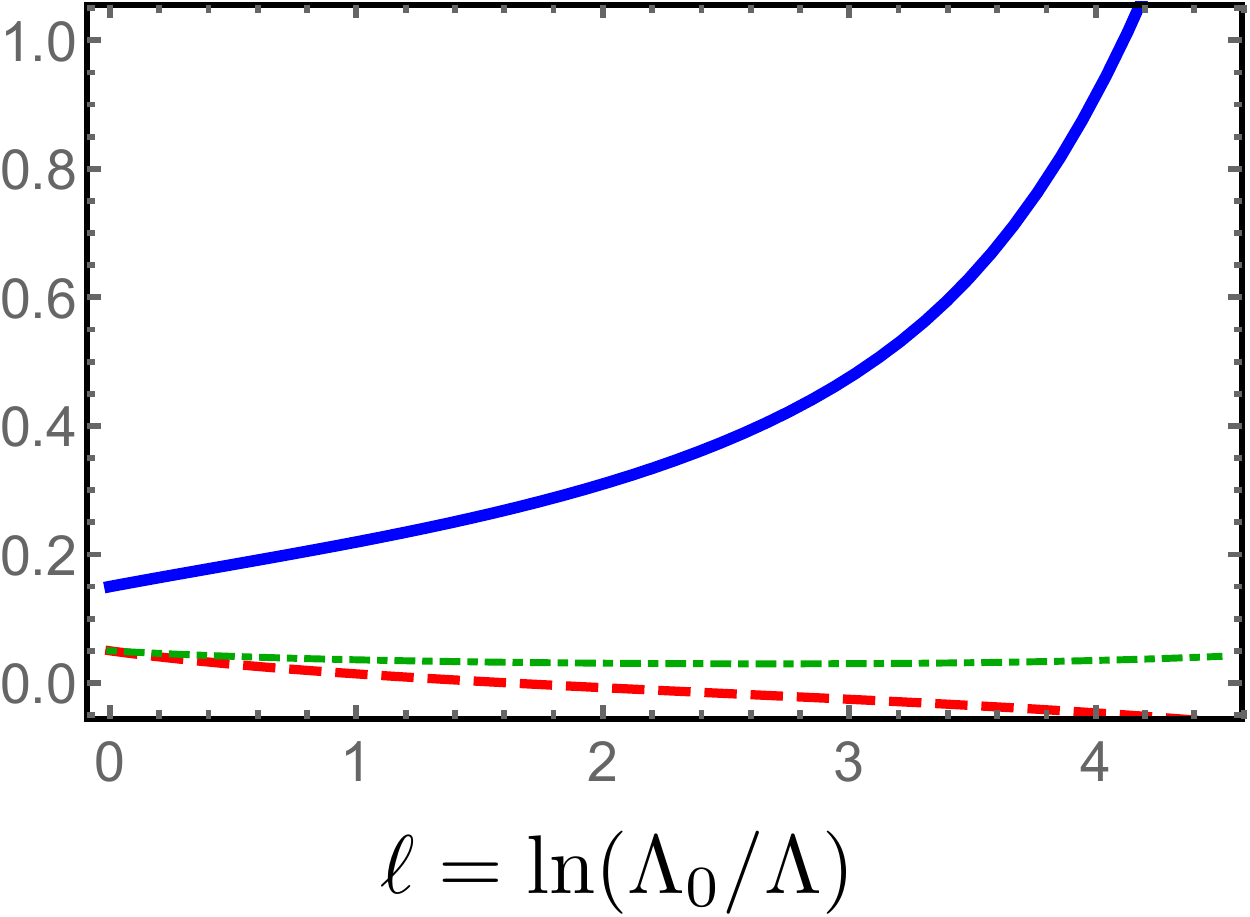}
  \caption{RG flow of marginal coupling constants $\lambda_+$ (solid line), $\lambda_-$ (dashed line) and $\gamma_{\text{tw}}$ (dot-dashed line) in the zigzag chain, according to Eq.  (\ref{eq:RG}). Here we set the initial conditions     $\tilde{\chi}(0)=0.1$,  $g(0)=\gamma_{\text{tw}}(0)=0.05$ and $\gamma_{\text{bs}}=-0.5$.   The nonuniversal prefactor is set to $C=1$. }
  \label{fig:rg1}
\end{figure}

Let us now discuss  the RG flow. First note that $\gamma_{\text{bs}}$ starts off with a bare value of order 1, but is marginally irrelevant  for $\gamma_{\text{bs}}<0$. To analyze the remaining equations, let us  assume $\tilde \chi >0$ without loss of generality. It is convenient to define \be \lambda_{\pm}=\tilde{\chi} \pm g.\ee
The combination of the $g$ and $\tilde\chi$ marginal perturbations can be written in the form\bea
\delta H_g+\delta H_{\chi}&=&2\pi v_s\lambda_+\int dx\,  \mathbf J_{e,R}\cdot \mathbf J_{o,L}\nonumber\\
&&+2\pi v_s\lambda_-\int dx\,  \mathbf J_{e,L}\cdot \mathbf J_{o,R}. \label{eq:Hg}
\eea
The RG equations  Eqs.~(\ref{eq:rg1})-(\ref{eq:rg3}) become\bea
\label{eq:RG}
\frac{d\lambda_+}{d\ell}&=&\lambda_+^2+2\pi ^2C^2\gamma_{\text{tw}}^2,\nonumber \\
\frac{d\lambda_-}{d\ell}&=&-\lambda_-^2-2\pi ^2C^2\gamma_{\text{tw}}^2,\nonumber \\
\frac{d\gamma_{\text{tw}}}{d\ell}&=&\left(\frac{\lambda_+-\lambda_-}2+\gamma_{\text{bs}}\right)\gamma_{\text{tw}}.
\eea
We are interested in the regime $\tilde{\chi}>g$, as follows from Eq.~(\ref{eq:kappa})   with  $ 1-\kappa \ll1 $. Physically, a sizeable three-spin interaction $\tilde \chi$ can be generated from virtual electron  hoppings in a Mott insulator in the vicinity of the metal-insulator transition~\cite{Motrunich2005,Bauer}. In this case $\lambda_+>0$  flows to strong coupling while $\lambda_-<0$ flows to zero.  Notice that  the twist operator contributes to  enhancing this trend.

Figure~\ref{fig:rg1}  shows a typical example of RG flow for a given choice of  bare coupling constants. The important point is  that $\lambda_+$ reaches strong coupling (\emph{i.e.} becomes of order 1) first. This behavior is   robust for a wide range of   initial values in the regime $\tilde \chi>g$. This implies that, in order to understand the properties of the low-energy fixed point,  we can analyze the effects of large $\lambda_+$ while dropping the other competing marginal operators. According to Eq. (\ref{eq:Hg}), the limit of coupling  $\lambda_+\to \infty$ gaps out    right movers in the even chain and left movers in the odd chain, but leaves the pair  of modes $\mathbf{J}_{e,L}$, $\mathbf{J}_{o,R}$ gapless. Furthermore, for small bare $\lambda_+(\ell=0)$ the gap in the $\mathbf{J}_{e,R}$, $\mathbf{J}_{o,L}$ pair (see Eq.~(\ref{Egap})) is exponentially small since the perturbation  is only marginally relevant. For the opposite chirality, $\tilde\chi<0$, the same picture holds   upon interchanging $\lambda_+ \leftrightarrow \lambda_-$ and reversing the pair of gapless modes.

The picture we have just described agrees with the low-energy spectrum for the integrable model discussed in Sec. \ref{sec:exact} if we identify the gapped modes with the excitations at $k\approx \pi$ in Fig. \ref{fig:spinons}. Moreover, the field theory analysis shows that  the 1D chiral spin liquid phase is generic (\emph{i.e.} does not depend on the fine tuning of coupling constants in Eq. (\ref{eq:kappa})) and is governed by the $\lambda_+$ operator, which gaps out a pair of left- and right-moving spin currents in neighboring chains.  This result  suggests a generalization to two dimensions, which we shall  discuss in the next section.

\section{Two-dimensional chiral spin liquid}
\label{se:2D}
\begin{figure}
\centering
\includegraphics*[width=.6\columnwidth]{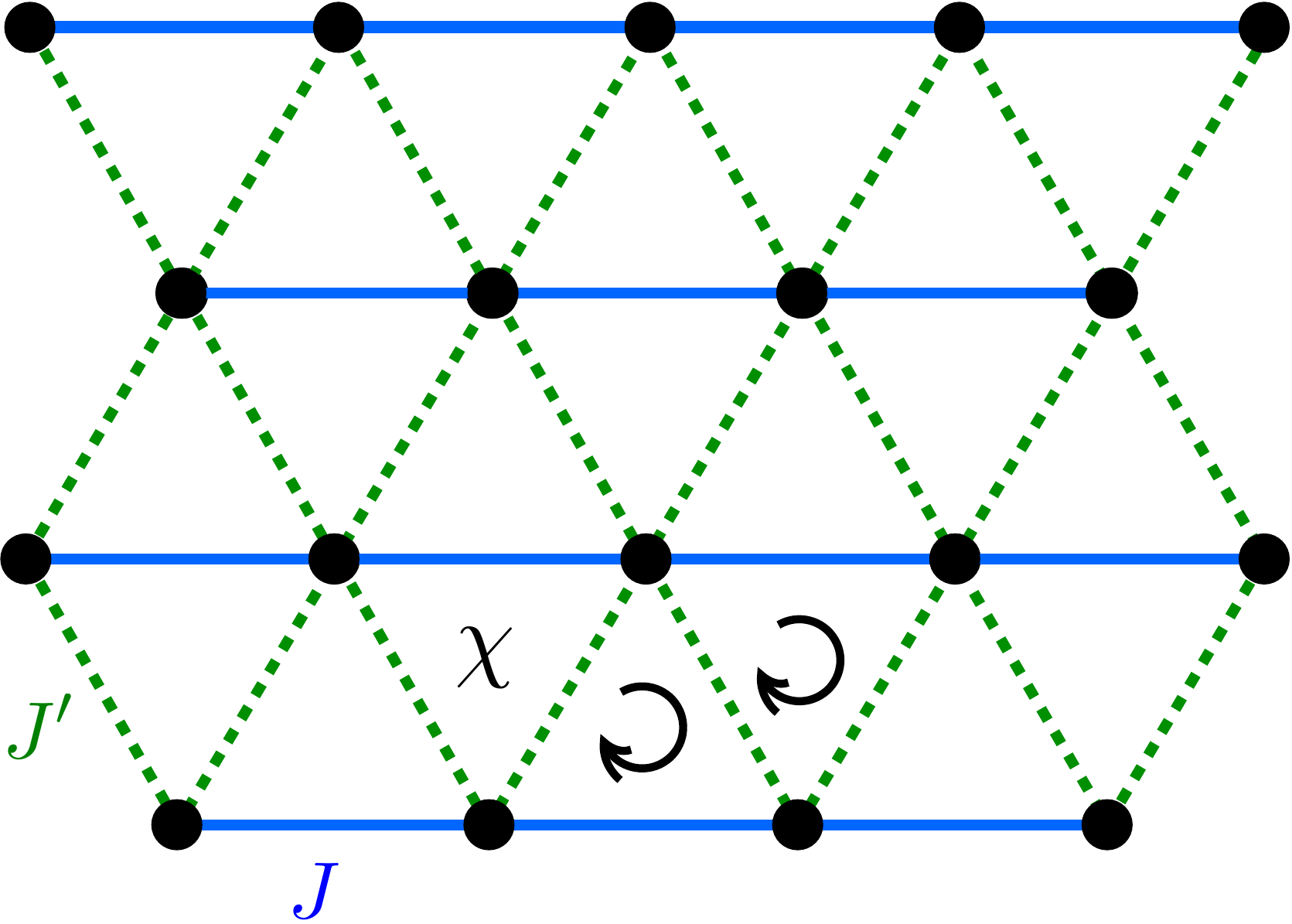}
\caption{Spatially anisotropic triangular lattice with exchange couplings $J\gg J^\prime $ and chiral three-spin interaction $\chi$. }
\label{fig:triangular}
\end{figure}

In the previous section we showed that the combination of the time-reversal-even interchain coupling $\delta H_g$ and time-reversal-odd $\delta H_\chi$ can lead to a phase   with gapless chiral modes  propagating in different legs of the zigzag chain. We now extend the argument to  frustrated 2D lattices built out of weakly coupled chains, such as the spatially anisotropic triangular lattice depicted in Fig.~\ref{fig:triangular}. The Hamiltonian is of the form $H = H_{J} + H_{\chi}$, where  \be
H_{J} = \sum_{ij} J_{ij} \mathbf{S}_i \cdot \mathbf{S}_j
\ee
contains the exchange couplings $J_{ij}=J$ for $i,j$ nearest-neighbor sites along horizontal links of the lattice (\emph{i.e.} within the same chain) and $J_{ij}=J^\prime\ll J$ for $i,j$ nearest-neighbor sites along diagonal links (\emph{i.e.} in neighboring chains). In addition, the Hamiltonian contains three-spin  operators with uniform chirality $\chi$,
\be
H_\chi = \frac{\chi}{2} \sum_{i , j, k \in \triangle } \mathbf{S}_i \cdot (\mathbf{S}_j \times \mathbf{S}_k ),\label{Hchitriang}
\ee
with the indices $i,j,k$ appearing clockwise in each elementary triangle.

The phase diagram of the spatially anisotropic triangular lattice has been studied in great detail in the   case of time-reversal-invariant interchain coupling~\cite{Starykh,ghamari}. The starting point is
a collection of $N$ decoupled  antiferromagnetic  Heisenberg  chains, each of which is described by  an SU(2)$_1$ WZW theory. The spin operator at position $x$ in chain $l$ is represented by \be
\mathbf S_{l}(x)\sim \mathbf J_{l,L}(x)+ \mathbf J_{l,R}(x)+(-1)^{x}\mathbf n_l(x),
\ee
where now $l=1,\dots,N$ is the chain (or leg) index. The  free Hamiltonian   for decoupled  chains reads
\be
\label{eq:H0N}
H_0=\sum_{l=1}^N\frac{2\pi v_s}{3}\int dx\, (\mathbf J_{l,R}^2+ \mathbf J_{l,L}^2).
\ee

Now consider the perturbations  that  are allowed by symmetry. Besides SU(2), translation $\mc L$ and time reversal $\mc T$, it is important to take into account the  $\mc P$ symmetry defined in Eq. (\ref{parity}), which for  an arbitrary number of chains takes  $x\to -x$ and \be
\mc P:\, \mathbf J_{l,R}\leftrightarrow  \mathbf J_{l,L},\, \,
\mathbf n_{l}\to (-1)^l \mathbf n_{l},\,\, \varepsilon_l\to(-1)^{l+1}\varepsilon_l.\label{parityNlegs}
\ee
As before, the intrachain backscattering process gives rise to the marginal perturbation \bea
\delta H_{\text{bs}}&=&2\pi v_s\gamma_{\text{bs}}\sum_{l}\int dx\, \mathbf J_{l,R}\cdot \mathbf J_{l,L},\eea
with $\gamma_{\text{bs}}\sim \mc O(1)$. The leading perturbations coupling first-neighbor chains are the generalizations of the  marginal operators discussed in Sec. \ref{se:fieldtheorzigzag}:\bea
\delta H_g&=&2\pi v_sg\sum_{l}\int dx\, (\mathbf J_{l,R}\cdot \mathbf J_{l+1,L}+R\leftrightarrow L),\label{HgNchains}\\
\delta H_{\text{tw}}&=&2\pi v_s\gamma_{\text{tw}}\sum_{l}(-1)^l\int dx\, \mathbf {n}_{l}\cdot \partial_x\mathbf n_{l+1},\label{HtwNchains}\\
\delta H_\chi&=&2\pi v_s\tilde \chi\sum_{l}\int dx\, (\mathbf J_{l,L}\cdot \mathbf J_{l+1,R}-R\leftrightarrow L),\label{HchiNchains}
\eea
where we have included the $\mc T$-breaking perturbation that stems from the three-spin interaction $\chi$. The perturbative RG equations for these marginal coupling constants are the same as in the 1D case,  Eqs. (\ref{eq:rg4}) through (\ref{eq:rg3}).

However, there is an important difference between the 1D and 2D cases. As discussed by Starykh and Balents~\cite{Starykh}, for $N>2$  there appear   two strongly   \emph{relevant} (dimension-1) perturbations that are allowed by symmetry and couple next-nearest-neighbohr chains:
\bea
\delta H_n &=&v_s  g_n \Lambda\sum_l \int dx\, \mathbf{n}_l \cdot \mathbf{n}_{l+2},\label{Hgn}\\
\delta H_\varepsilon &=&v_s   g_\varepsilon  \Lambda\sum_l \int dx\,  \varepsilon_l \varepsilon_{l+2},\label{Hgeps}
\eea
where $g_n$ and $g_\varepsilon$ are dimensionless. Note that these operators respect the $\mc P\circ \mc T$ symmetry with $\mc P$ defined in Eq. (\ref{parityNlegs}). These are the only allowed relevant perturbations even in our case where  $\mc P$ and $\mc T$ are separately  broken by the spin chirality operator. The RG equations for $g_n$ and $g_\varepsilon$ read
\bea
\frac{d g_n}{d\ell}&=&\left(1-\frac{\gamma_{\text{bs}}}2\right)g_n,\\
\frac{dg_\varepsilon}{d\ell}&=&\left(1+\frac32\gamma_{\text{bs}}\right)g_{\varepsilon},
\eea
where we have included the correction due to the $\mc O(1)$ marginal coupling $\gamma_{\text{bs}}$. As argued in Ref. [\onlinecite{Starykh}], the effect of $\gamma_{\text{bs}}<0$ is to enhance the growth of $g_n$; for $g_n >0$, this  favors an instability towards a collinear antiferromagnetic  phase in which the $\mathbf n_l(x)$ fields are pinned. By contrast, for $\gamma_{\text{bs}}>0$ (which  can happen if one adds a  sufficiently large intrachain next-nearest-neighbor exchange coupling)  a dimerization instability driven by the $\varepsilon_l(x)$ fields becomes dominant.

\begin{figure}
\centering
\includegraphics*[width=.8\columnwidth]{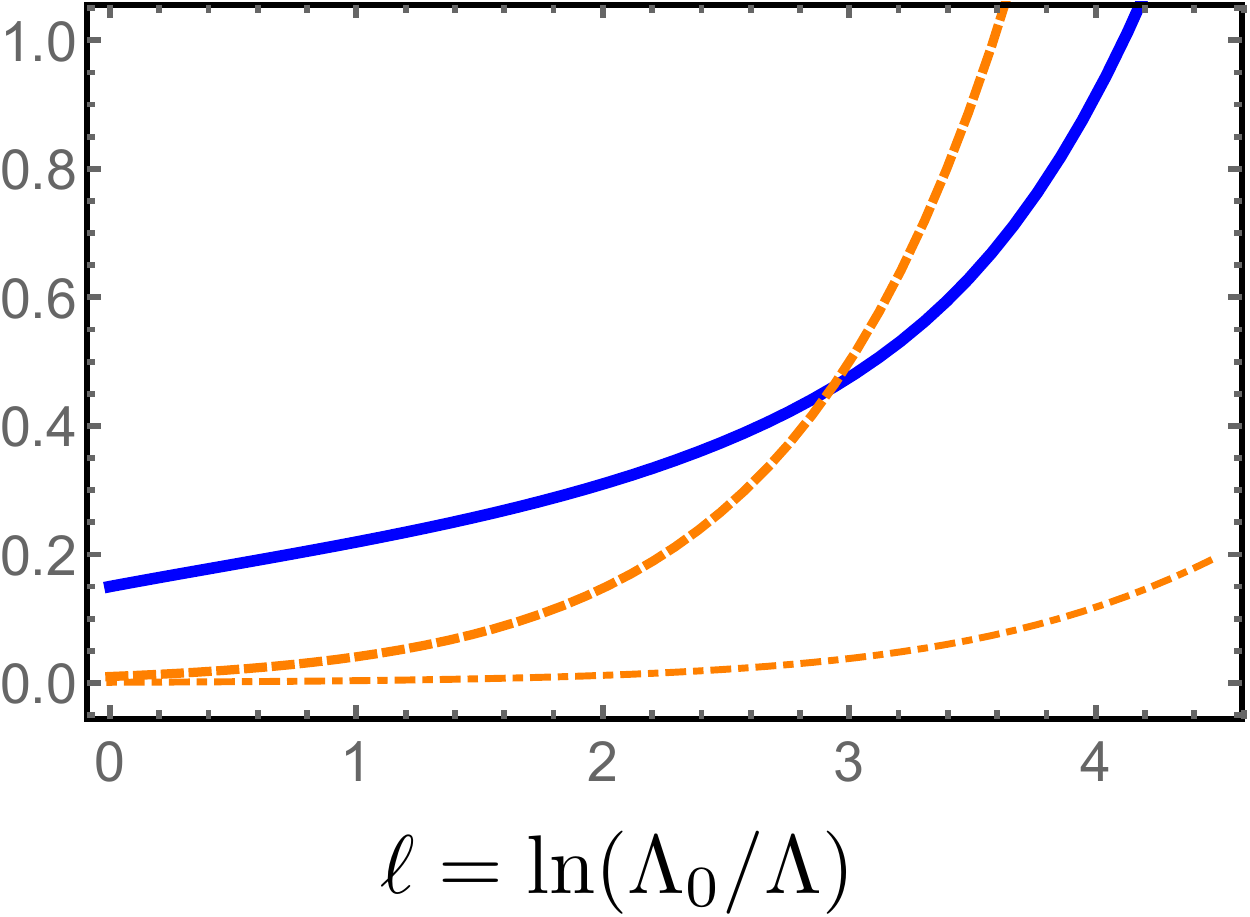}
\caption{RG flow for the marginal coupling constant $\lambda_+$ (solid line) and relevant coupling constant $g_n$. For the latter, we show two curves corresponding to two different initial conditions: $g_n(0)=0.01$ (dashed line) and $g_n(0)=0.001$ (dot-dashed line). The other initial values used in this plot are $\tilde \chi(0)=0.1$, $g=\gamma_{\text{tw}}=0.05$ and $\gamma_{\text{bs}}=-0.5$.  The dashed line represents a case in which the relevant coupling $g_n$ starts off smaller than $\lambda_+$ but reaches strong coupling first, driving the system into a phase with long-range magnetic order. The dotted line represents the case in which $\lambda_+$ reaches strong coupling first and leads to the CSL phase. }
\label{fig:rggood}
\end{figure}

Naively, one would expect either one of the relevant operators $g_n$ or $g_\varepsilon$ to  overtake the marginal couplings   in Eqs. (\ref{HgNchains}-\ref{HchiNchains}) and  govern the low-energy physics, leading to more conventional, long-range-ordered phases. However, the  fate of the system also depends on the bare values of the coupling constants, and cases in which  a marginal operator reaches strong coupling before relevant ones have been discussed  in the literature~\cite{Egger,Hikihara10,ghamari}.
In the following we shall assume $\gamma_{\text{bs}}<0$ and focus our discussion on $g_n$ as the most relevant operator that competes with $\lambda_+$. For instance, Fig. \ref{fig:rggood} shows the RG flow for $g_n$ and $\lambda_+$ for two different values of the bare $g_n(\ell=0)\ll \lambda_+(\ell=0)$.  We see that which operator reaches strong coupling first  is a quantitative question, whose answer is sensitive to the precise initial conditions of the RG flow. Nevertheless, in the following we shall attempt to make statements about the typical qualitative behavior depending on how $g_n(0)$ scales with the interchain  couplings $J^\prime,\chi\ll J$.

Let us then estimate the magnitude of the bare $g_n(0)$ in Eq. (\ref{Hgn}). Our  original lattice model   does not contain direct coupling between next-nearest-neighbor chains. However, since the operator is allowed by symmetry, we expect it to be generated by the RG flow at higher orders in $J^\prime,\chi$.  For the triangular lattice without the three-spin interaction, Starykh and Balents~\cite{Starykh} found that $g_n>0$  is generated during the initial stages of  the RG only at order $(J^\prime/J)^4$. On the other hand,  it has been suggested~\cite{ghamari} that a ferromagnetic  $g_n(0)<0$  is expected from fluctuations at short length scales at order $(J^\prime/J)^2$. The latter  is more consistent with existing numerical results which have not observed the collinear antiferromagnetic phase, but rather incommensurate spiral order~\cite{ghamari,weichselbaum}.

In our case, the chiral spin interaction provides another source of the relevant coupling between second-neighbor chains. Assuming that the initial value $g_n(0)$ is set by fluctuations at short length scales, we can argue that in the triangular lattice $g_n$ is generated at order $\chi^2$. The argument is based on a perturbative calculation in a real-space picture. We proceed along the lines of Refs. [\onlinecite{stoudenmire,Starykhkagomme}]. Let $H_0$ denote the Hamiltonian of decoupled chains and $|0\rangle$ be the corresponding ground state with energy $E_0$. We regard $H_\chi$ in Eq. (\ref{Hchitriang}) as a perturbation to $H_0$. We define the projectors $P=|0\rangle \langle 0|$  and $Q=1-P$, and write $|\Psi_0\rangle =P|\Psi\rangle$ for the projection of an arbitrary state $|\Psi\rangle$. One can then derive an eigenvalue equation for $|\Psi_0\rangle$ in the form $H_{\text{eff}}|\Psi_0\rangle=E|\Psi_0\rangle$, with an effective Hamiltonian given by\cite{stoudenmire,Starykhkagomme} \be
H_{\text{eff}}=H_0+PH_\chi (1-RQH_\chi)^{-1}RH_\chi,
\ee
where $R=(E-H_0)^{-1}$ is the resolvent operator. To second order in perturbation theory, we can approximate $R\approx R_0=(E_0-H_0)^{-1}$ and $H_{\text{eff}}\approx H_0+V_{\text{eff}}^{(2)}$ with the effective interaction\be
V_{\text{eff}}^{(2)}=PH_\chi R_0H_\chi.\label{effectiveint}
\ee

\begin{figure}
\centering
\includegraphics*[width=.8\columnwidth]{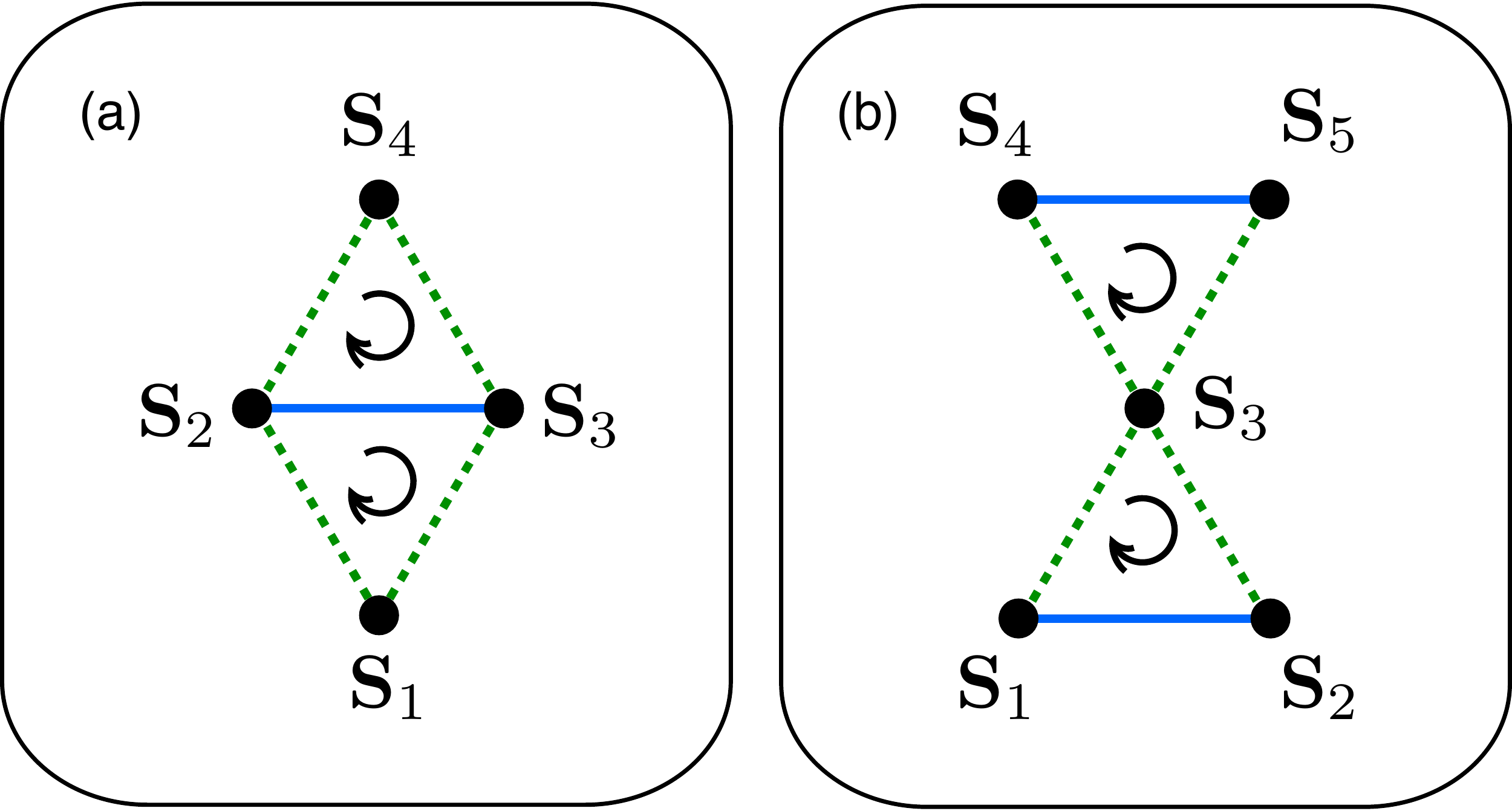}
\caption{(a) Edge-sharing triangles; (b) corner-sharing triangles.  Only in the first case is the relevant coupling $g_n$ generated at order $\chi^2$.}
\label{fig:edgecorner}
\end{figure}

Consider now the edge-sharing triangles represented in Fig. \ref{fig:edgecorner}(a). The spins $\mathbf S_1$ in the lower chain and $\mathbf S_4$ in the upper chain both interact with  the spins $\mathbf S_2$ and $\mathbf S_3$ in the middle chain via the chiral three-spin interaction.  The exchange coupling between   $\mathbf S_1$ and $\mathbf S_4$ can be generated at second order in $H_\chi$ using Eq. (\ref{effectiveint}) and  projecting out  the middle chain. We find\be
V_{\text{eff}}^{(2)} \sim \chi^2 G^{+-}\mathbf S_1\cdot \mathbf S_4,\label{coupleS1S4}
\ee
where \be
G^{+-}=-i\int_0^{\infty}dt\, \langle T^{+-}_{2,3}(t)T^{+-}_{2,3}(0)\rangle
\ee
is the zero-frequency retarded Green's function for the two-spin operator\be
T^{+-}_{2,3}=\frac{i}{2}(S_{2}^+S_{3}^--S_{2}^-S_{3}^+).
\ee
Note that $T^{+-}_{2,3}$ is equivalent to  the $z$ component of the spin current flowing between sites $2$ and $3$ (the choice of the component is arbitrary due to SU(2) symmetry). Alternatively, via Kramers-Kronig relations~\cite{Starykhkagomme} $G^{+-}$ can be expressed in terms of the dynamical structure factor for the operator $T^{+-}_{i,j}$ involving nearest-neighbor sites $i,j$ in a single Heisenberg chain. Note that $T^{+-}_{i,j}$ can be viewed as the   antisymmetric part (a vector related to the spin current operator) of the two-spin tensor operator $T^{ab}_{i,j}$, $a,b\in \{x,y,z\}$. To our knowledge, only the dynamical structure factor   for the scalar part of this tensor, $\sum_aT^{aa}_{i,j}\sim \mathbf S_i\cdot \mathbf S_j$, has been calculated by exact methods~\cite{caux}. In any case, the main point is that by taking the staggered parts of the spin operators  in Eq. (\ref{coupleS1S4}), we obtain the interaction $V_{\text{eff}}^{(2)} \sim g_n\mathbf n_l\cdot \mathbf n_{l+2}$ (where $l$ and $l+2$ denote the chains that contain $\mathbf S_1$ and $\mathbf S_4$, respectively) with $g_n$ of order $\chi^2$.

From the above discussion, we conclude that for the triangular lattice with $\chi>J^\prime$ we should in general expect $g_n(0)\sim \mc O(\chi/J)^2$. Since the relevant operator is  generated already at this level of perturbation theory and $g_n(\ell)$ grows exponentially fast with $\ell$, regardless of the sign of $g_n(0)$, it seems rather unlikely that the marginal coupling $\lambda_+$ will reach strong coupling first and lead to the CSL phase in the triangular lattice, and instead one expects a more conventional order~\cite{privatecom}.

More propitious  conditions for stabilizing the CSL   are found in lattices where the chiral three-spin interaction is confined to \emph{corner-sharing} triangles. Consider a set of  spins connected as in Fig. \ref{fig:edgecorner}(b). Repeating  the perturbative analysis at short-length scales, we verify  that  projecting out the spin $\mathbf S_3$ in the intermediate corner gives rise to a coupling $\propto \sum_{a,b}(S_1^aS_2^b-S_1^bS_2^a)(S_4^aS_5^b-S_4^bS_5^a)$. Importantly, the operator $S_1^aS_2^b-S_1^bS_2^a$ involving the spins in the lower chain  is odd under link parity but even under time reversal. Thus, taking the continuum limit cannot produce $\mathbf n_l(x)$ in the chain containing $\mathbf S_1$ and $\mathbf S_2$. Therefore, this $\mc O(\chi^2)$ perturbative calculation does not generate the relevant coupling $g_n\mathbf n_l\cdot \mathbf n_{l+2}$. At the same time, second order in $J^\prime$   generates an operator $\propto (\mathbf S_1+ \mathbf S_2)\cdot(\mathbf S_4+ \mathbf S_5)$. The operator in the lower chain $\mathbf S_1+ \mathbf S_2$ is odd under time reversal but even under link parity; thus, in the continuum limit it does not generate $\mathbf n_l$ at order $(J^\prime)^2$ either. While our argument is based on a lattice  picture,   the same conclusion can be reached by integrating out fast modes in the initial steps of the RG in the continuum limit (as done \emph{e.g.} in Ref. [\onlinecite{Starykh}]). Note also that the same arguments can be used to rule out  $g_\varepsilon(0)$ at order $\chi^2,(J^\prime)^2$. We conclude that the corner-sharing triangular geometry has a higher degree of frustration in the sense that the relevant coupling  between second-neighbor chains  is pushed to higher orders in $\chi$ and $J^\prime$. In this case we expect $g_n\sim \mc O(\chi/J)^4$.

\begin{figure}
\centering
\includegraphics*[width=.6\columnwidth]{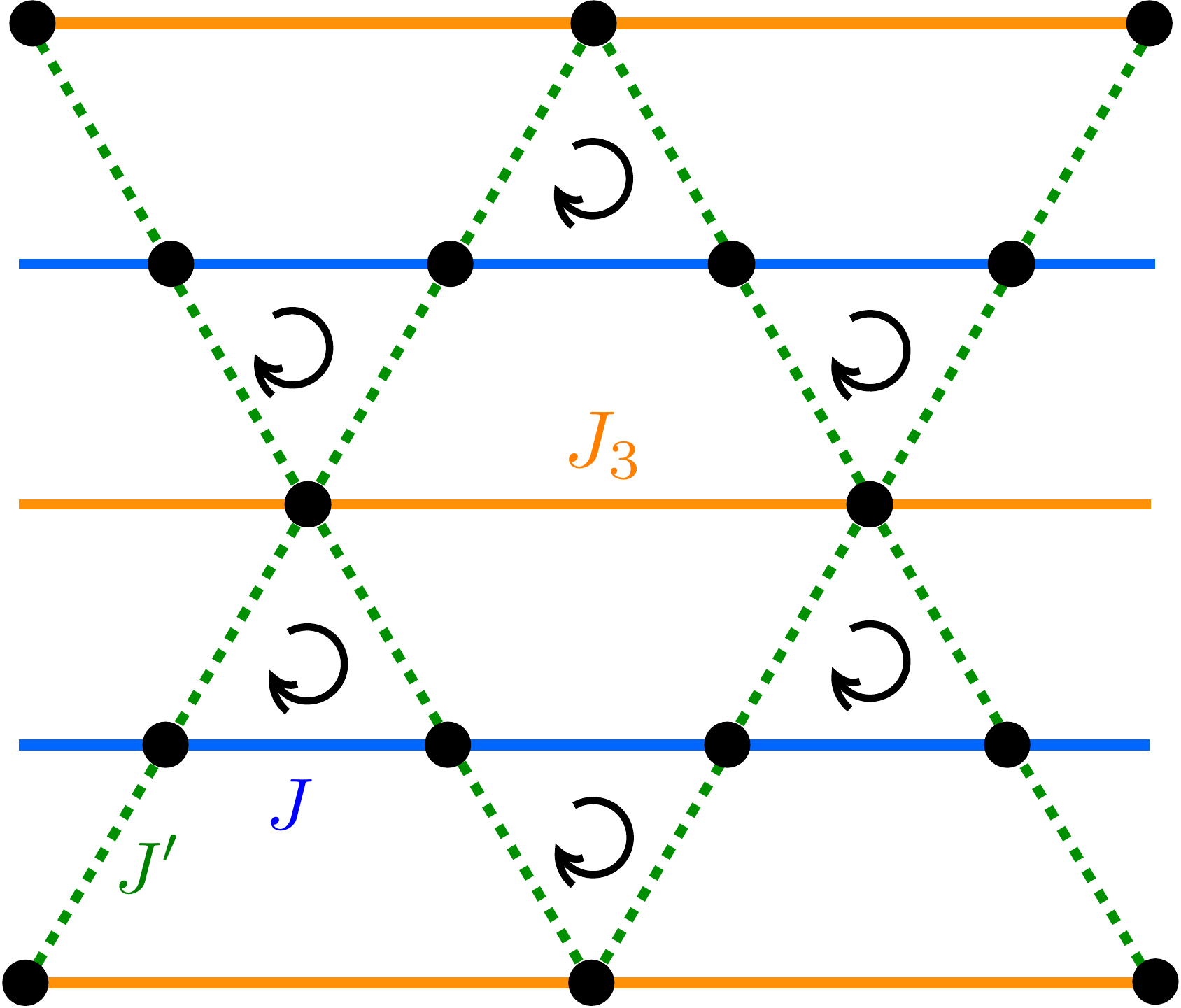}
\caption{Spatially anisotropic 2D lattice with   corner-sharing triangles, derived from the kagome lattice by adding  a third-neighbor coupling $J_3$  along the horizontal lines inside the hexagons.}
\label{fig:kagome}
\end{figure}

In Fig. \ref{fig:kagome} we show an example of an anisotropic 2D lattice constructed by coupling chains with only corner-sharing triangles.  This lattice differs from the anisotropic   kagome lattice\cite{Starykhkagomme} by  an additional exchange coupling $J_3$ between third neighbors in the horizontal direction across the hexagons. Moreover, our model includes the three-spin interaction in each triangle.  For $J^\prime,\chi\ll J_3\lesssim J$,  the starting point for our analysis is that all spins on this lattice belong to a Heisenberg chain. But in this case there are two types of chains with different site densities, which we call \emph{dense chains} (with coupling $J$) and \emph{dilute chains} (with coupling $J_3$). It is worth mentioning that another motivation for considering the $J_3$ coupling is that the extended kagome lattice studied in recent DMRG simulations~\cite{hesheng,Gong,GongZhuBalents} required relatively  large (spatially isotropic) second- and third-neighbor couplings in order to stabilize the CSL phase.

We can directly apply the  perturbative  argument about corner-sharing triangles to show  that between second-neighbor dense chains the relevant $g_n$ coupling is not  generated at order $\chi^2,(J^\prime)^2$. Meanwhile, the sites on the dilute chains form a triangular sublattice; thus,  the $g_n,g_\varepsilon$ couplings between nearest dilute chains are ruled out by symmetry.  Therefore,  $g_n$ and $g_\varepsilon$ must be fourth order in $\chi,J^\prime$. In this case, we expect the scenario represented by the dot-dashed line in Fig. \ref{fig:rggood}, \emph{i.e.} the initial values of $g_n,g_\varepsilon$ are so small that the marginally relevant  coupling  $\lambda_+$ reaches strong coupling first and gaps out pairs of $R$ and $L$ modes in neighboring chains.

In the next section we shall study the properties of  a 2D state  dominated by the operator $ \lambda_+\sum_l\mathbf J_{l,R}\cdot \mathbf J_{l+1,L}$, showing that this is indeed the Kalmeyer-Laughlin CSL.

\section{Topological Properties}
\label{se:properties}
Building on the results of the previous section, in this part we will assume that the perturbation\be
\delta H_{+}=2\pi v_s\lambda_+\sum_{l=1}^{N-1} \int dx\, \mathbf J_{l,L}\cdot \mathbf J_{l+1,R}\label{eq:HgN}
\ee
is the leading relevant operator and gaps out     pairs of  chiral currents $\mathbf{J}_{l,L} ,   \mathbf{J}_{l+1,R}$ in first-neighbor chains. This is similar to the case of the zigzag  chain discussed in Sec. \ref{se:fieldtheorzigzag}, except that now  the modes that remain   gapless  are   spatially separated \emph{edge states}, composed of  the right-moving spin mode  in the $l=1$ chain and the left-moving spin mode  in the $l=N$ chain. This is a concrete realization of the idea of merging triangular puddles to form a 2D topological phase in the network model perspective~\cite{Bauer}.   Thus, the low-energy theory of each edge is described by a \emph{chiral} WZW SU(2)$_1$ model.  Note that there are still symmetry-allowed relevant perturbations that can couple these edge modes, but their coupling constants decrease exponentially with $N$ and the effect can be neglected in the 2D limit.

Since the edge states do not carry charge, the Hall conductivity vanishes. However, at low temperature $T$ the chiral edge modes carry  an energy current flowing counterclockwise
around the edge, given by $J_Q=\frac{\pi c}{12}T^2$, where $c=1$ is the   central charge in this case~\cite{kanefisher,cappelli}.

This critical theory of the edge is consistent with the properties of the Kalmeyer-Laughlin CSL state.  To completely characterize a 2D topological state, one needs to account for the correct bulk physics in addition to the edge physics. Below we will discuss the bulk quasiparticles and show that they correspond to  spin-$1/2$ anyons. The unambiguous  signature of a topological state is its degeneracy on the torus. We will demonstrate that the state dominated by Eq.~(\ref{eq:HgN}) is doubly degenerate when placed on a torus. This is directly linked with the exchange statistics of these spin-1/2 quasiparticles, implying that they are anyons with statistical phase $\theta=\pi/2$~\cite{WenNiu}.

\subsection{Bulk quasiparticle excitations}
\label{se:QP}
Quasiparticle (QP) excitations can be constructed using a semiclassical picture that follows from the strong coupling limit of the interchain coupling Eq.~(\ref{eq:HgN}). This strong coupling picture is easily understood using the methods of Kane, Mukhopadhyay, and Lubensky~\cite{kane_02} for the FQHE. In this description, each 1D chain consists of $L$ and $R$ bosonic   fields, $\varphi_{l,L/R}$,  subject to a cosine perturbation of the form $\cos[A(\varphi_{l,R} -\varphi_{ l+1,L} )]$, with some constant $A$ that depends on the scaling dimension of local operators. In the strong coupling limit, the field difference $\varphi_{l,R} -\varphi_{ l+1,L} $ is localized   in one of the minima of the cosine potential, and  QP excitations are solitonic solutions corresponding to jumps between adjacent  minima.

It is instructive to begin our construction of QPs using the above  method, even though  it does not   display the SU(2) symmetry of the Kalmeyer-Laughlin CSL. This symmetry gives rise to  extra degeneracies and   implies that the QPs must transform under an irreducible representation of SU(2). In fact, this restricts the possible fractionalization of quantum numbers in systems with SU(2) symmetry. As described below,  non-Abelian bosonization  is the natural language to construct the bulk QPs with   explicit symmetry properties.

Let us separate the longitudinal and transverse parts of the  interchain interaction,  $\protect{\mathbf{J}_{l,L} \cdot   \mathbf{J}_{l+1,R} }$ $= \mathcal{O}^z_{l+  \frac{1}{2}} +\mathcal{O}^{xy}_{l+  \frac{1}{2}} $, where\bea \mathcal{O}^z_{l+  \frac{1}{2}}&=& J^z_{ l,L}  J^z_{ l+1,R},\\
\mathcal{O}^{xy}_{l+  \frac{1}{2}}&=& \frac{1}{2}(J^+_{ l,L}  J^-_{ l+1,R} +\text{h.c.}).\eea
Here $l+1/2$  represents the link between chains $l$ and $l+1$. We use   Eq.~(\ref{Jabelian}) to represent the   components of  spin currents in terms of bosonic fields.  The transverse part of the interchain coupling  yields
\bea
\mc O^{xy}_{l+\frac12}=\frac{1}{4\pi^2 }\cos[\sqrt{4 \pi}(\varphi_{l,L} -\varphi_{ l+1,R} )].\label{transversecosine}
\eea
Upon flowing to strong coupling, this operator pins the field difference $\varphi_{l,L} -\varphi_{ l+1,R} $ to the minimum of the cosine potential; hence in the ground state of the gapped phase
\be
\label{pinnedValue}
\sqrt{4 \pi}(\varphi_{l,L} -\varphi_{l+1,R} ) = 2 \pi n + \pi,~~~n \in \mathbb{N}.
 \ee
Consider an excitation in which  the argument of the cosine in Eq. (\ref{transversecosine}) jumps by $\pm 2 \pi$ over some finite region in space (the size of the QP, which depends on the energy gap).
The total spin $\Delta S^z$ accumulated over that region is
\bea
\Delta S^z&=&\sum_l \int dx\, (J^z_{l,L}+J^z_{l,R})\nonumber\\
&=&\frac{1}{\sqrt{4 \pi}} \sum_l \int dx\, \partial_x (\varphi_{l,L} - \varphi_{l+1,R} )\nonumber\\
&=& \pm  \frac{1}{2} .
\eea
Thus, QPs have eigenvalues of $S^z$  equal to  $\pm 1/2$.

In the above discussion  we have   ignored the longitudinal  operator $\mathcal{O}_{l+\frac12}^{z}$. We now check that this was legitimate. Summing over chain index, we can write
\bea
\sum_l \mathcal{O}^{z}_{l+  \frac{1}{2}}&=&\frac{1}{8\pi } \sum_l [ (\partial_x \varphi_{l,L}-\partial_x \varphi_{l+1,R})^2  \nonumber \\
&&-(\partial_x \varphi_{l,R})^2-(\partial_x \varphi_{l,L})^2  ].\label{Ozint}
\eea
In the SU(2) symmetric case, the longitudinal and transverse parts of the marginally relevant operator flow together to strong coupling. As the transverse part $\mc O^{xy}_{l+\frac12}$ locks the difference $\varphi_{l,L}- \varphi_{l+1,R}$ in the low-energy limit, the first term in Eq. (\ref{Ozint}) vanishes.  The remaining terms are equivalent to   a renormalization of the  spin velocity which does not change the qualitative  features  of the RG flow. Thus,  the operator $\mathcal{O}_{l+\frac12}^{z}$ does not affect the strong-coupling picture of pinning the fields as in Eq.~(\ref{pinnedValue}).

We can also write down    the QP creation operator. The latter  should create a $\pm 2 \pi$ kink in the field difference $\sqrt{4 \pi}(\varphi_{l,L} -\varphi_{ l+1,R} )$ at link $l+1/2$. Using the commutation relations in Eq. (\ref{eq:commut}), we can easily identify  this with the vertex operators    $  e^{\pm i \sqrt{\pi} \varphi_{l,L}(x)}$ (or equivalently   $  e^{\mp i \sqrt{\pi} \varphi_{l+1,R}(x)}$, since the chiral bosons are locked together in the ground state). Taking the SU(2) symmetry into account, we recognize that the vertex operators that create QPs with $\Delta S^z=+1/2$ or  $\Delta S^z=-1/2$ are the two components of the chiral spinor in the  WZW  model (\emph{cf}. Eq. (\ref{gAbelian})):
\bea
\label{eq:QP}\Psi^{QP}_{l+\frac{1}{2}} (x)&\propto  & g_{l,L}(x) =  \left(\begin{array}{c}
      e^{-i \sqrt{\pi} \varphi_{l,L}(x)} \\
      e^{i \sqrt{\pi} \varphi_{l,L}(x)}
     \end{array}
    \right) .
\eea
The spinor structure of the QP operator makes it explicit that it forms a spin-1/2 representation of the SU(2) spin-rotational  symmetry. This also implies that ``particle'' ($\Delta S^z=+1/2$) and ``hole'' ($\Delta S^z=-1/2$) excitations can be continuously rotated into each other, which of course is only possible because the QPs are charge neutral.  Furthermore, local physical operators can only create pairs of QPs. For instance, the dimerization operator $\varepsilon_l(x)\propto \text{tr}[g_l(x)]$, which is an SU(2) scalar, creates spin singlet   excitations, whereas the staggered magnetization $\mathbf n_l(x)\propto \text{tr}[\boldsymbol \sigma g_l(x)]$ creates triplet   excitations.

\subsection{Topological degeneracy}
\label{se:GSdegeneracy}
The imprint of topological order is a ground state degeneracy which is sensitive to the topology of the space~\cite{WenNiu}. To establish  the ground state  degeneracy on the torus,  it suffices to  find two operators $U_x , U_y$ that commute with the Hamiltonian but not with each other.  This implies that  the ground-state manifold  must form  a representation of the algebra obeyed  by $U_x$ and $U_y$   which is necessarily multidimensional. More generally, the number of such operators grows with the genus of the surface~\cite{WenNiu}. In the FQHE case, this algebra is of the form~\cite{WenNiu} $U_x U_y=U_y U_x e^{2 \pi i/m}$ with $m$ integer corresponding to the filling factor $\nu=1/m$. If we work in the basis of the $U_x$  operator and label a ground state by $| x \rangle$, such that   $U_x |x \rangle = x | x \rangle$, then the state $U_y |x \rangle$ is also  a ground state but has $U_x$ eigenvalue of $x e^{2 \pi i/m} \ne x$. Only after applying $U_y$   $m$ times do we return to the same value of $x$, implying that there are at least $m$ different ground states.  Recently the coupled-wire approach was used to construct the $U_x$ and $U_y$ operators in the FQHE~\cite{Sagi}.  In the following we demonstrate   this structure for  $m=2$ in our case of a CSL    governed    by  the interaction in  Eq.~(\ref{eq:HgN}).

One operator that obviously commutes with any lattice spin Hamiltonian is $e^{i 2 \pi  S^z_{l_0}}$, where $S^z_{l_0}$ is the $z$ component of the total spin  operator in an arbitrary chain $l=l_0$. This operator is either equal to  $1$ for an even number of spins in the chain or to $-1$ for an   odd number. Upon coupling with other chains, the total spin of the $l_0$-th chain  can  only change by an integer; thus,   $e^{i 2 \pi  S^z_{l_0}}$ stays invariant. As a result,  $[e^{i 2 \pi  S^z_{l_0}},H] = 0$.

Focusing  on the low-energy  theory, we write  our Hamiltonian   simply as  $H=H_0+\delta H_+$, with $H_0$ in Eq. (\ref{eq:H0N}) and $\delta H_+$  in Eq.~(\ref{eq:HgN}). Now consider the operator
\be
\label{eq:X}
U_x =   e^{i 2\pi  \int dx J^z_{l_0,L}(x) }.
\ee
This is almost the same as above, except that  it only involves  the total spin in the left-moving  chiral sector  of the $l_0$-th chain, \emph{i.e.}  $U_x =   e^{2\pi i    S^z_{l_0,L }}$ with  $S^z_{l_0,L } =\int dx\, J^z_{l_0,L}(x)$. In abelian bosonization notation,
 \be
 \label{eq:XAbelian}
U_x = e^{i \sqrt{ \pi} \int dx \partial_x \varphi_{l_0,L}(x) }.
 \ee
One can check explicitly that $[H , U_x]=0$ since $H$ is written in terms of $\mathbf{J}_{l,L}$ and $\mathbf{J}_{l,R}$.
Indeed, using the commutation relations Eq.~(\ref{eq:commut}) and the expressions in Eq. (\ref{Jabelian}), we verify that $J^\pm_{l_0,L}$ changes $S^z_{l_0,L } $ by $\pm 1$, so it does not affect $U_x$.

We can then work in the basis of  $U_x$. We   label states by the eigenvalue of $S^z_{l_0,L }$, which can be split  into integer and fractional parts, $| S^z_{l_0,L }  \rangle  = | \mathcal{N}+f \rangle$, where $\mathcal{N} \in \mathbb{N}$ and $f = S^z_{l_0,L } {\text{ mod}} ~1$. Since the eigenvalues of $U_x$ do not depend on $\mathcal{N}$ but only on $f$, we denote these states only by the fractional part $|f \rangle$.

If $U_x$ commutes with all components of  $\mathbf J_{l,\alpha}$, how can one find a physical operator that does not commute with $U_x$? The answer is that  the theory allows for  physical  operators which do not appear in $H$ but change the spin in a chiral sector of a given chain   by a fractional value. In the WZW model, one such operator is   the staggered magnetization  $\mathbf{n}_l(x) = {\rm{tr}}[g_l(x) \boldsymbol{\sigma}]$. In abelian bosonization, we can write the $z$ component of $\mathbf n_l$ as \bea
n^z_l(x)&\propto& \sin\{\sqrt\pi[\varphi_{l,L}(x)-\varphi_{l,R}(x)]\}\nonumber\\
&=&\frac1{2i}\,e^{i\sqrt\pi[\varphi_{l,L}(x)-\varphi_{l,R}(x)]}+\text{h.c.}.
\eea
This is a dimension-$1/2$ vertex operator, clearly distinct from the dimension-1 chiral currents. Using the commutation relations Eq.~(\ref{eq:commut}), we can verify that   $n_l^z(x)$ changes the eigenvalue of $S^z_{l_0,L }$ by $\pm 1/2$. Thus, the fractional part $f$ that labels the eigenstates changes by $1/2$, \emph{i.e.}  $n^z_l(x)$ switches between  the two  sectors with $|f=0\rangle $ and $|f=1/2\rangle$.

We can write down a linear combination of $n_l^z(x)$ and $\varepsilon_l(x)\propto \cos\{\sqrt\pi[\varphi_{l,L}(x)-\varphi_{l,R}(x)]\}$   to construct an operator that commutes with $H$ but does not commute with $U_x$. Consider
\bea
U_y&=&\prod_l  e^{i\sqrt\pi[\varphi_{l,L}(x_0)-\varphi_{l,R}(x_0)]},
\eea
where $x_0$ is an arbitrary position in the chain direction. Note that $U_y$  does not change   the fractional part of the total $S^z_{l}$ for  any given chain, but  it changes the fractional part of each chiral sector $S^z_{l,\alpha}$ separately. The algebra of $U_x,U_y$ can be obtained using $e^A e^B = e^B e^A e^{[A,B]}$ and   Eq.~(\ref{eq:commut}). Since
  \be
  [i \sqrt{\pi} \int dx\, \partial_x \varphi_{l_0,L}(x) ,\sqrt{\pi} \sum_{l} \varphi_{l,L } (x_0)]= i \pi ,
  \ee
 we find \be
  U_x U_y = - U_y U_x.\label{anticommute}
 \ee
 Rather than demanding $[H,U_y]=0$, it is actually sufficient to show that $U_y$ does not change the energy starting from any ground state. As long as the system is closed into a torus in the $y$ direction and the fields are locked in the ground-state manifold according to  Eq. (\ref{pinnedValue}), we can write
\bea
U_y &= &\prod_le^{i\sqrt\pi[\varphi_{l+1,L}(x_0)-\varphi_{l,L}(x_0)]}.\label{Uylockedfields}\eea
But in the phase dominated by the relevant perturbation Eq.~(\ref{eq:HgN}) the   difference appearing in the exponential in Eq. (\ref{Uylockedfields}) is just a constant; thus, this operator acts as a constant in the ground-state manifold.  It   follows from  Eq. (\ref{anticommute}) that $|f=1/2\rangle \propto U_y|f=0\rangle$ is a ground state orthogonal to $|f=0\rangle$, and our CSL state has the same topological degeneracy on the torus as the $\nu=1/2$ FQHE state.

\begin{figure}
\centering
\includegraphics*[width=.55\columnwidth]{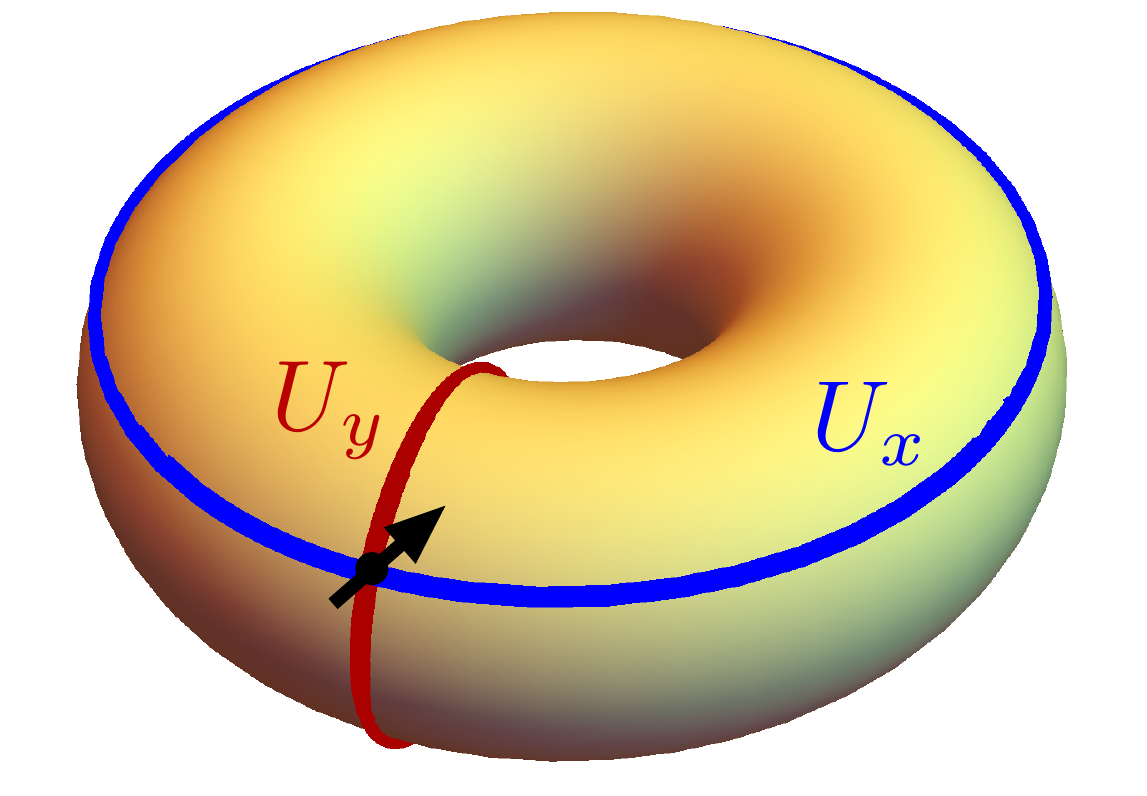}
\caption{Chiral spin liquid on a torus. The operators $U_x$ and $U_y$ transport  a spin-$1/2$ quasiparticle along the $x$ direction (parallel to the chains) and $y$ direction (perpendicular to the chains), respectively.}
\label{fig:spinonontorus}
\end{figure}

The operators $U_x$ and $U_y$ have the physical interpretation  of transporting QPs around the torus in the $x$ or $y$ directions, respectively~\cite{wen_book,Sagi} (see Fig. \ref{fig:spinonontorus}).  Moreover, it was   demonstrated in Ref. [\onlinecite{WenNiu}] that for anyons with statistical phase $\theta$ the commutation relation between $U_x$ and $U_y$ is $ U_x U_y = e^{2 i \theta} U_y U_x$. In our case, this implies that the spin-1/2 QPs in the CSL are anyons with statistical phase $\theta =\pi/2$ (\emph{i.e.} semions).

Having established the correspondence between the edge states, quasiparticle properties  and topological degeneracy, we conclude that the state described in this section is equivalent to the Kalmeyer-Laughlin CSL.

\section{Summary and Outlook}
\label{sec:conclusion}
We presented a coupled-chain construction of chiral spin liquids. It was first applied as a long-wavelength description to an exactly solvable lattice model in one dimension, and then   generalized to two dimensions. In the latter case our formulation assumed a dominant relevant interchain coupling given by Eq.~(\ref{eq:HgN}), which stems from a chiral three-spin interaction. This formulation yields all the universal properties of the Kalmeyer-Laughlin CLS state, suggesting their equivalence.

The   energy gap $\Delta$ of the CSL state is exponentially small in the chiral interchain coupling $\tilde \chi$, \emph{i.e.}   $\Delta \propto J e^{-\frac{1}{\tilde\chi}}$, reflecting the nonperturbative nature of the present approach. On the other hand,  the  smallness of the gap raises the concern  that there could be other competing instabilities  that might  dominate the low-energy physics. By means of a careful renormalization group analysis, we found  that  the interaction  responsible for stabilizing the CSL can  reach  strong coupling before other relevant perturbations with parametrically small coefficients. This scenario  is expected in the kagome lattice rather than the triangular lattice. While we have considered the spatially anisotropic limit of weakly coupled Heisenberg chains, we expect the phase discussed here to be  adiabatically connected with the CSL observed in recent numerical work on the isotropic extended kagome lattice~\cite{Bauer,hesheng,Gong,GongZhuBalents,HuZhuBecca,Wietek}.

We have not explored here the possibility of spontaneous breaking of time reversal symmetry giving rise to the spin chirality order parameter. This possibility can in principle be investigated    using a mean-field decoupling of time-reversal-invariant interchain interactions. Another interesting question is whether one can use the present coupled-chain approach to construct a Z$_2$ quantum spin liquid~\cite{wenz2}, which was shown to be stabilized on the kagome lattice~\cite{YanHuseWhite}.  Many additional possibilities are offered by the current approach, including generalizations to more exotic chiral spin liquids, for example SU(N) CSLs with proposed realizations for ultracold fermionic alkaline earth atoms~\cite{Hermele09}; other exotic states can be obtained by  starting from SU(2)$_k$ WZW models, which  can be realized in higher-$S$ spin chains~\cite{AffleckHaldane,thomale12a,Greiter09,Scharfenberger11}. We leave the development of  these ideas for future study.

\acknowledgements
We thank E. Bettelheim, Y. Gefen, D. B. Gutman,  Y. Oreg, E. Sagi, and R. A. Santos for illuminating discussions.  This work was supported by ISF and Marie Curie CIG grants (E.S.) and by CNPq (R.G.P.).

\emph{Note added:} In the final stages of this work, we became aware that a similar idea is being pursued by T. Meng, T. Neupert, M. Greiter, and R. Thomale~\cite{MengThomale}.

\bibliography{chiralspinliquid}

\end{document}